\documentclass[nofootinbib,prd,twocolumn,showpacs,showkeys,preprintnumbers]{revtex4-1}
\usepackage{hyperref,amssymb,amsmath,mathrsfs,bm,graphicx}
\usepackage{array}

\usepackage[dvipsnames]{xcolor}

\usepackage[maxfloats=256]{morefloats}
\begin{document}

\title {Geodesic analysis and steady accretion on a traversable wormhole}

\author{A. Rueda}
\affiliation{Departamento de F\'isica, Colegio de Ciencias e Ingenier\'ia, Universidad San Francisco de Quito,  Quito 170901, Ecuador.\\}

\author{E. Contreras }
\email{econtreras@usfq.edu.ec}
\affiliation{Departamento de F\'isica, Colegio de Ciencias e Ingenier\'ia, Universidad San Francisco de Quito,  Quito 170901, Ecuador.\\}


\begin{abstract}
In this work, we analyze the behavior of light and matter as they pass near and through a traversable wormhole throat. In particular, we study the trajectories of massive and massless particles and the dust accretion around a traversable wormhole previously reported in Eur. Phys. J. C \textbf{82} (2022) no.7, 605.
For massive particles, we integrate the trajectory equation for ingoing and outgoing geodesics and classify the orbits of particles scattered by the wormhole in accordance with their asymptotic behavior far from the throat. We represent all the time--like trajectories in an embedding surface where it is shown explicitly the trajectories of i) particles that deviate from the throat and remain in the same universe, ii) particles that traverse the wormhole to another universe, and iii) particles that get trapped in the wormhole in unstable circular orbits. For the massless particles, we numerically integrate the trajectory equation to show the ray-tracing around the wormhole specifying the particles that traverse the wormhole and those that are only deviated by the throat. For the study of accretion, we consider the steady and spherically symmetric accretion of dust. Our results show that the wormhole parameters can significantly affect the behavior of light and matter near the wormhole. Some comparisons with the behavior of matter around black holes are made.
\end{abstract}
\maketitle
\section{Introduction}

Wormholes are hypothetical structures in spacetime that could act as shortcuts between two distant points in the universe, potentially allowing for faster-than-light travel and time travel. The concept of wormholes can be traced back to Einstein and Rosen's work \cite{einstein1935particle} where they introduced the concept of a ``bridge'' between distant points in the spacetime. However, the original Einstein-Rosen bridge is not traversable, meaning that it does not allow for the passage of matter or information from one end to the other.

The concept of a traversable wormhole was first introduced by Morris and Thorne \cite{morris1988wormholes} attracting a lot of attention in the community. Since then, the topic has remained an active area of research leading to some interesting results in the context of general relativity and beyond (the list of reference is extense but you can see \cite{Mustafa:2023ojf, Rehman:2023hro, Errehymy:2023rnd,Errehymy:2023rsm,Rahaman:2023tkm,Mustafa:2023hlu,Mustafa:2023kqt,Dias:2023pdx,Sengupta:2023ysx,Mustafa:2023pko,Estrada:2023pny,Ali:2022hho,Bzowski:2022kgf,Tayde:2022vbn,Capozziello:2022zoz,Zubair:2023abs,Bhattacharya:2023imf, Hassan:2022ibc,Tello-Ortiz:2022hyf,Tello-Ortiz:2021kxg,Maldacena:2020sxe, Konoplya:2021hsm,Mustafa:2022fxn, Mehdizadeh:2020nrw,Simpson:2019oft, Garattini:2019ivd,Cataldo:2017ard,Bronnikov:2003gx,Lemos:2003jb,Kuhfittig:2015cea,Debnath:2013iya}, and reference there in, for the last ten years). Nevertheless, to maintain traversable wormhole stability and prevent its collapse, the existence of exotic matter with negative energy density is mandatory \cite{visser1995lorentzian, Lobo2017} (although this requirement can be avoided in the context of modified theories of gravity  \cite{Nilton:2022cho, Nilton:2022hrp}). However, exotic matter is purely theoretical and has not yet been observed in nature but its existence is a necessary assumption for wormholes to be viable in current theories. Nevertheless, although there is no experimental evidence to support the existence of exotic matter, some proposals as the Casimir effect appear as the only mechanism to produce an artificial source of traversable wormholes, namely, the Camimir energy \cite{Garattini:2019ivd}. For these reasons, all the developments on traversable wormholes are focused on their potential observation as they could behave as black hole mimickers.

In recent years, some efforts have been made to construct suitable (theoretical) wormholes based on
the analysis of the frequencies of the quasinormal modes and the quantifier of exotic matter \cite{KimThorne1991, HochbergVisser1997, Lobo2005, HarkoLoboMak2008, KonoplyaZhidenko2016, CardosoFranzinPani2016}. These studies, not only allow us to explore the response of the geometry under perturbations but they help to discriminate between traversable wormholes and black holes based on the pattern of their oscillatory behavior. Another important tool in the characterization of traversable wormholes (and distinguishing them from black holes) is the geodesic analysis near and around their throat which can reveal unique features of the geometry \cite{AnchordoquiTorres1997,guerrero_olmo_rubiera-garcia_gómez_2022,Chandrasekhar:1985kt,mishra_chakraborty_2018,taylor_2014,Olmo:2021piq,Perlick2004,Ovgun2018,JusufiOvgun2019,cardoso2020light,FernandesJusufi2021}. Similarly, the study of accretion on traversable wormholes can exhibit specific patterns which are not observed in the case of black holes. By analyzing these distinct features, we can gain a deeper understanding of the nature of the compact object in question, and potentially identify the presence of a traversable wormhole as opposed to a black hole \cite{Kardashev2013,AzregAinouEksi2015,Sokoliuk:2022owk,mauryaBH}.

In this work, we perform a comprehensive geodesic analysis \cite{Chandrasekhar:1985kt} of both timelike and null geodesics in the context of traversable wormholes based on the solution reported in \cite{RuedaAvalosContreras2022}. In particular, we integrate the equation of the trajectory of particles approaching the wormhole throat and analyze to what extent the parameters of the solution affect the path of both, particles that remain in the same universe and particles that traverse the throat and appear in the other universe. Besides, we explore the steady spherically symmetric (Bondi--Michel--like) accretion \cite{Bondi:1952ni,michel1972} of pressureless fluid (dust) on the traversable wormholes and compare these results with the obtained from other geometries in the literature.

The remainder of this paper is organized as follows: Section \ref{wormhole} presents a brief overview of the static and spherically symmetric traversable wormhole metric. Section \ref{geodesic} details our geodesic analysis, with a focus on timelike and null geodesics. In Section \ref{accretion}, we discuss the accretion of dust around the wormhole, and in the last section the conclusion of the work and final remarks.

\section{Wormhole model}\label{wormhole}
In this section, we summarize the main aspects related to static and spherically symmetric wormholes and the specific model we use in this work. We begin by considering a spherically symmetric line element as follows:
\begin{eqnarray}\label{metric}
ds^{2}=-e^{2\phi} dt^2 +d r^{2}/(1-b/r)+r^{2}(d\theta^{2}+\sin^{2}\theta d\phi^{2}),\nonumber\\
\end{eqnarray}
where $\phi=\phi(r)$ and $b=b(r)$ depend on the radial coordinate only and denote the redshift and shape functions, respectively. The line element (\ref{metric}), is the standard parameterization of a traversable wormhole because it allows to implement the conditions that ensure the desired geometry in a natural way. For example, in order for the wormhole to be traversable, we must demand that the redshift function is finite everywhere, which is equivalent to demand $e^{2\phi}\ne0$. Besides, the particular form of the $g_{rr}$ component of the metric, allows for connecting the shape function $b(r)$ with an embedding function after comparing the metric with $dt=0$ and $\theta=\pi/2$ with a general metric in cylindrical coordinates (see the details below).

Given that (\ref{metric}) is a solution to Einstein's field equations,
\begin{eqnarray}\label{EFE}
R_{\mu\nu}-\frac{1}{2}g_{\mu\nu}R=\kappa T_{\mu\nu},
\end{eqnarray}
with $\kappa=8\pi G/c^{4}$ (assuming $c=G=1$ in this work), and sourced by $T^{\mu}_{\nu}=diag(-\rho,p_{r},p_{t},p_{t})$, the only non-vanishing equations read
\begin{eqnarray}
\rho&=&\frac{1}{8\pi}\frac{b'}{r^{2}}\label{rho}\\
p_{r}&=&-\frac{1}{8\pi}\left[\frac{b}{r^{3}}-2\left(1-\frac{b}{r}\right)\frac{\phi'}{r}\right]\label{pr}\\
p_{t}&=&\frac{1}{8\pi}\left(1-\frac{b}{r}\right)
\bigg[\phi''+(\phi')^{2}-\frac{b'r-b}{2r^{2}(1-b/r)}\phi'\nonumber\label{pt}\\
&&-\frac{b' r-b}{2r^{3}(1-b/r)}+\frac{\phi'}{r}\bigg],
\end{eqnarray}
where $\rho$, $p_{r}$ and $p_{t}$ stand for the energy density, the radial and the transverse pressure of the source, respectively.
Next, we describe the primary properties of a traversable wormhole through its embedding in three-dimensional Euclidean space. Owing to the symmetry of the solution, we can assume $\theta=\pi/2$ without loss of generality. For a constant time $t$, the line element can be expressed as

\begin{eqnarray}\label{emb1}
ds^{2}=\frac{dr^{2}}{1-b/r}+r^{2}d\phi^{2}.
\end{eqnarray}

Note that Eq. (\ref{emb1}) represents the line element of a surface so it can be embedded in $\mathbf{R}^{3}$ if we compare it with a metric in cylindrical coordinates $(r,\phi,z)$, namely
\begin{eqnarray}
ds^{2}=dz^{2}+dr^{2}+r^{2}d\phi^{2}.
\end{eqnarray}
Now, by assuming that  $z$ depends on the radial coordinate, we obtain
\begin{eqnarray}
dz=\frac{dz}{dr}dr,
\end{eqnarray}
which results in
\begin{eqnarray}\label{emb2}
ds^{2}=\left[1+\left(\frac{dz}{dr}\right)^{2}\right]dr^{2}
+r^{2}d\phi^{2}.
\end{eqnarray}

Comparing (\ref{emb1}) and (\ref{emb2}), we deduce
\begin{eqnarray}\label{emb3}
\frac{dz}{dr}=\pm 
\left(\frac{r}{b}-1\right)^{-1/2},
\end{eqnarray}
where it is evident that $b>0$ for $r\in[r_{0},\infty)$. It is worth mentioning that, after solving (\ref{emb3}) we obtain the so-called embedding function $z(r)$ which allows us to ``visualize'' the traversable wormhole structure). Several important remarks should be made at this point: (i) The wormhole geometry must have a minimum radius, causing $dz/dr\to\infty$ as $r\to b_{0}$ (which occurs when $b=r$). Consequently, the existence of a minimum radius necessitates that $b=b_{0}$ at $r=b_{0}$.
(ii) We demand an asymptotically flat solution, implying both $b/r\to0$ (from which $dz/dr\to0$) and $\phi\to0$ as $r\to\infty$.
(iii) The smoothness of the geometry is ensured if the embedding surface flares out at or near the throat which is represented by the flaring-out condition
\begin{eqnarray}\label{foc}
\frac{b-b'r}{2b^{2}}>0.
\end{eqnarray}
It is worth noticing that the flaring-out condition (\ref{foc}) results in a violation of the null energy condition (NEC) which can be demonstrated by defining the quantity
\begin{eqnarray}\label{exp}
\xi=-\frac{p_{r}+\rho}{|\rho|}=\frac{b/r-b'-2(r-b)\phi'}{|b'|},
\end{eqnarray}
which contains the main ingredients appearing in the expression for the NEC for an anisotropic fluid, namely $\rho+p_{r}$. Now, Eq. (\ref{exp}) can be rewritten as
\begin{eqnarray}
\xi=\frac{2b^{2}}{r|b'|}\frac{d^{2}r}{dz^{2}}
-2(r-b)\frac{\phi'}{|b'|}.
\end{eqnarray}
Note that, as $(r-b)\to0$ at the throat, we find
\begin{eqnarray}
\xi=\frac{2b^{2}}{r|b'|}\frac{d^{2}r}{dz^{2}}>0
\end{eqnarray}
so that
\begin{eqnarray}
\xi=-\frac{p_{r}+\rho}{|\rho|}>0.
\end{eqnarray}

If $\rho>0$, the above condition implies $p_{r}<0$, which indicates that $T^{1}_{1}$ should be interpreted as a tension. Furthermore, defining $\tau=-p_{r}$, the flaring-out condition results in
\begin{eqnarray}
\tau-\rho>0,
\end{eqnarray}
which means that, for this exotic matter, the throat tension must be greater than the total energy density, violating the NEC as mentioned earlier.

From a technical standpoint, constructing traversable wormholes involves solving the system (\ref{rho})-(\ref{pt}), which consists of three equations with five unknowns, namely $\{\phi,b,\rho,p_{r},p_{t}$\}. In this work, we shall use the model developed in Ref. \cite{RuedaAvalosContreras2022} where the authors considered a null redshift function $\Phi=0$ and an analytic embedding function
\begin{equation}\label{z}
  z (r)=\sqrt {\log\left (a + \left (\frac {c r} {r_0} + d \right)^2 \right)},  
\end{equation}
where $r_0$ is the wormhole throat and $a$, $c$, and $d$ are free parameters (please do not confuse the free parameter $c$ with the speed of light involved in Eq. (\ref{EFE}) which we have set to one). It is worth mentioning that such an embedding function was proposed in order to ensure the flaring-out condition under certain restrictions of the parameters involved. Besides, this function leads to a suitable shape function in the sense that the solution is asymptotically flat. Indeed,
the shape function resulting from (\ref{emb3}) is
\begin{equation}
       b(r)=\frac{c^2\zeta^2 r}{c^2\zeta^2+r_0^2\left( a +\zeta^2 \right)^2\log\left( a+\zeta^2\right)}
\end{equation}
where
\begin{eqnarray}
\zeta=\left( \frac {c r} {r_0} +d \right).
\end{eqnarray}
and 
\begin{equation}
    a = 1 - c^2 - 2 c d - d^2,
\end{equation}
imposed by the flaring-out condition.

In Ref. \cite{RuedaAvalosContreras2022}, the authors not only constructed the wormhole geometry but they analyzed its stability through the quasinormal modes by using the
Wentzel-Kramers-Brillouin (WKB) semi-analytical approach (for an incomplete list see \cite{Konoplya:2023moy, Konoplya:2022tvv, Churilova:2021nnc, Konoplya:2020jgt, Panotopoulos:2020mii, Rincon:2020pne}, and references therein). In this work, instead, we explore the behavior of test particles around the wormholes through geodesic analysis and steady accretion as we shall see in what follows.

\section{Geodesic equations}\label{geodesic}
In this section, we will deepen the analysis of geodesic equations in the context of traversable wormholes. Geodesic equations provide a powerful tool to study the motion of test particles and light in a given spacetime, revealing crucial information about the underlying geometry and potential observational signatures. The study of geodesics in traversable wormholes is particularly important as it helps us to understand the fundamental differences between wormholes and other compact objects, such as black holes or neutron stars, for example. 

We will begin by deriving the geodesic equations for a static, spherically symmetric traversable wormhole spacetime. We will then explore the solutions to these equations and discuss their physical implications. Particular attention will be paid to the behavior of timelike and null geodesics near the wormhole's throat and how it is influenced when the free parameters $(c,d)$ change. 

\begin{figure*}[hbt!] 
\centering
 \includegraphics[width=0.3\textwidth]{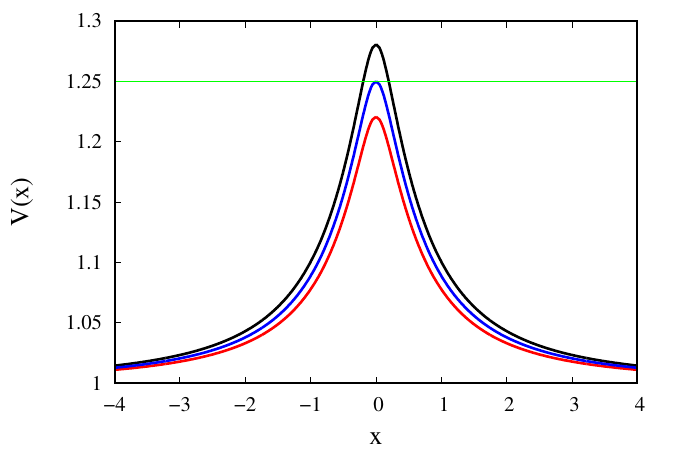}\
\includegraphics[width=0.3\textwidth]{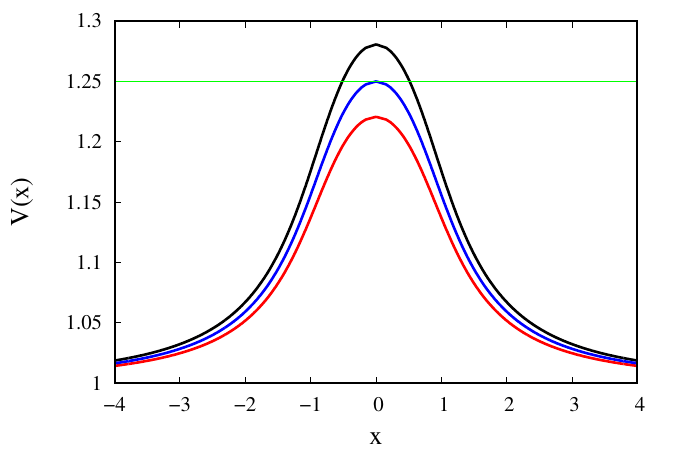}\
\includegraphics[width=0.3\textwidth]{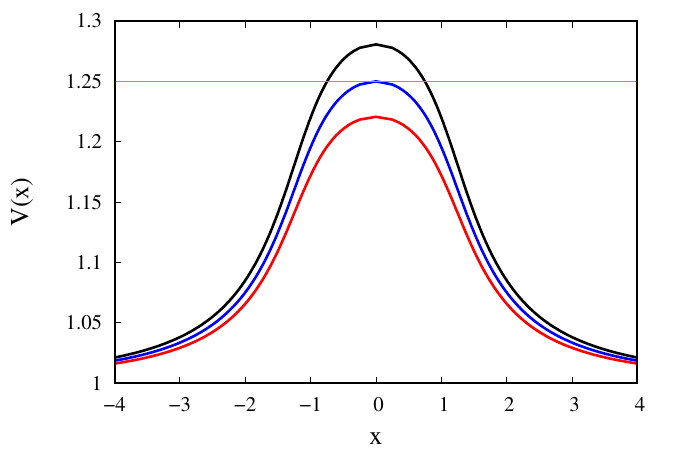}\
\caption{\label{tlp}
Effective potential of time-like geodesics as a function of the tortoise coordinate. This is for $c = 0.4$, $d = 0.2$ (left), $c = 1.2$, $d = 0.6$ (center) and $c = 1.6$, $d = 1.8$ (right). The green line represents energy, the red curve is the potential for $L = L_c - \delta$, and the blue curve is the potential for $L = L_c + \delta$, for $\delta = 0.05$. 
}
\end{figure*}

The equatorial ($\theta=\pi/2$) geodesic equations in static and spherically symmetric space-times read \cite{Chandrasekhar:1985kt,mishra_chakraborty_2018,taylor_2014}
 \begin{align}
&\dot{t}=E e^{-2\Phi(r)}\\
&\dot{\phi}=\frac{L}{r^2}\\
&\dot{r}^2=\left( 1-\frac{b(r)}{r} \right)\left( E^2 e^{-2\Phi(r)} -\epsilon -\frac{L^2}{r^2}\right),
\end{align}
with $L$ and $E$, the angular momentum and energy of the particle, respectively. Besides, $\dot{}$ represents the derivative with respect to an affine parameter and $\epsilon = 1,0$ corresponds to timelike and null geodesics. 
In the particular case of traversable wormholes with vanishing redshift, the equations are given by
 \begin{align}
     \dot{t}&=E\\
     \dot{\phi}&=\frac{L}{r^2}\label{phidot}\\
     \dot{r}^2&=\left( 1-\frac{b(r)}{r} \right)\left( E^2 -\epsilon -\frac{L^2}{r^2}\right),\label{rdot}
 \end{align}
Now, it is convenient to introduce the so-called tortoise coordinate as it allows to connect both universes of the solution by avoiding coordinate divergences. In terms of the tortoise coordinate $x$ given by 
\begin{eqnarray}\label{tort}
x(r)=\int_{r_0}^r \frac{1}{\sqrt{1-b(r')/r'}}dr',
\end{eqnarray}
Eq. (\ref{rdot}) reads
\begin{align}\label{eqgeo}
     \dot{x}^2 = E^2 - V^2(x),
 \end{align}
where $V(x)$ is the effective potential expressed as a function of $x$, defined as
 \begin{equation}\label{effpot}
     V(x)=\sqrt{\epsilon+\frac{L^2}{r(x)^2}}.
 \end{equation}
It is worth mentioning that Eq. (\ref{eqgeo} only depends on $x$ so it allows) us to analyze an effective 1D problem of the orbits (as usually done in this context) by evaluating the permitted orbits given the shape of the potential $V(x)$ and the energy of the particle. More precisely, we can use \eqref{eqgeo} to classify orbits. In this case, the potential is bell-shaped with its maximum at $x=0$ so we will classify the orbits by taking energies near such a maximum.  If $E > V(x=0)$ the particle goes through the throat and moves on to infinity, if $E = V(x=0)$ the particle follows an unstable circular orbit around the throat, and if $E < V(x=0)$ the particle cannot traverse the throat and moves on to infinity in the same universe. The previous classification can be performed in terms of the critical angular momentum obtained from Eq. (\ref{effpot}) after imposing the condition
 $E = V(0)$, which leads to
 \begin{equation}
L_{c}=\sqrt{E^{2}-\epsilon}r_{0},
 \end{equation}
that allows us to write the angular momentum of the particles near the throat of the wormhole as $L = L_c \pm \delta$, for $\delta > 0$. In terms of this definition, we note that if $L = L_c$ the particle describes a circular orbit around the throat. However, if $L = L_c + \delta$ the particle remains in the same universe, and if  $L = L_c - \delta$  the particle goes through the throat and travels to the infinity of the other universe. In Fig. \ref{tlp} is shown the effective potential as a function of $x$ for timelike geodesics with energy specified by the green line. The effective potentials correspond to 
the cases $L>L_{c}$ (red line), $L<L_{c}$ (black line) and $L=L_{c}$ (blue line). 

It is worth emphasizing that Fig. \ref{tlp} is just a one--dimensional representation of what really occurs so, in order to gain insight about the trajectory described by the particle we proceed to obtain the equation for the trajectory of the particle. By combining  (\ref{phidot}) and (\ref{rdot}) we arrive at
\begin{equation}
    \frac{d\phi}{dr} = \pm \frac{L}{r^2 \sqrt{\left( 1-\frac{b(r)}{r}\right)\left(\epsilon+E^2 -\frac{L^2}{r^2} \right)}},
\end{equation}
where $+$ and $-$ stand for ingoing and outgoing geodesics, respectively. Assuming $r_i $ and $\phi_i $ as the initial position of the particle in polar coordinates, we obtain that the integral for the ingoing geodesic is 
\begin{equation}\label{geo1}
    \phi_{in} (r) =  \int_{r_i}^{r} \frac{L}{r'^2 \sqrt{\left( 1-\frac{b(r')}{r'}\right)\left(\epsilon+E^2 -\frac{L^2}{r'^2} \right)}} dr' + \phi_i
\end{equation}
calculated  from $(r_i, \phi_i)$ to $(r_{min},\phi_{in}(r_{min}))$, where $x_{min}$ is given by $\dot{x_{min}}=0$, in which the particles follow an outgoing geodesic.
The integral for the out-going geodesic with the new initial condition $\phi_i'=\phi_{in}(r_{min})$ reads
\begin{equation}\label{geo2}
    \phi_{out} (r) = - \int_{r_{min}}^{r} \frac{L}{r'^2 \sqrt{\left( 1-\frac{b(r')}{r'}\right)\left(\epsilon+E^2 -\frac{L^2}{r'^2} \right)}}dr' + \phi_i'
\end{equation}

\begin{figure*}[hbt!] 
\centering
\includegraphics[width=0.3\textwidth]{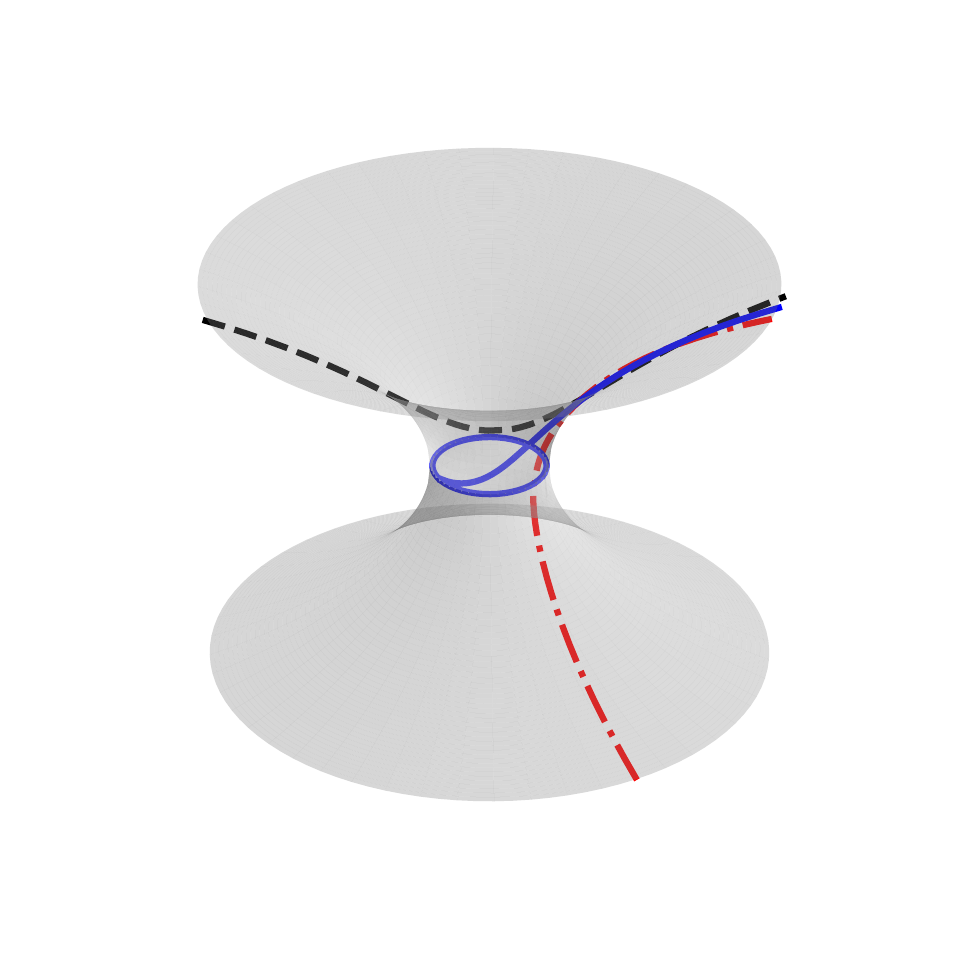}\
\includegraphics[width=0.3\textwidth]{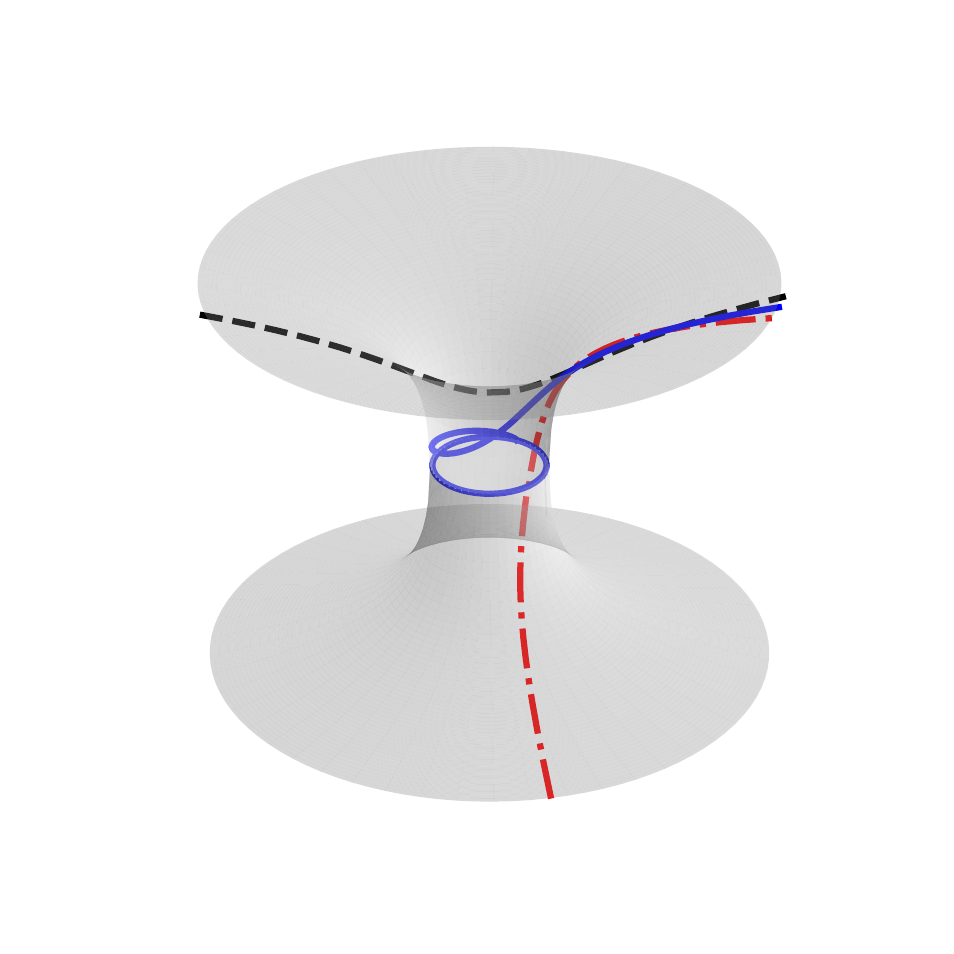}\
\includegraphics[width=0.3\textwidth]{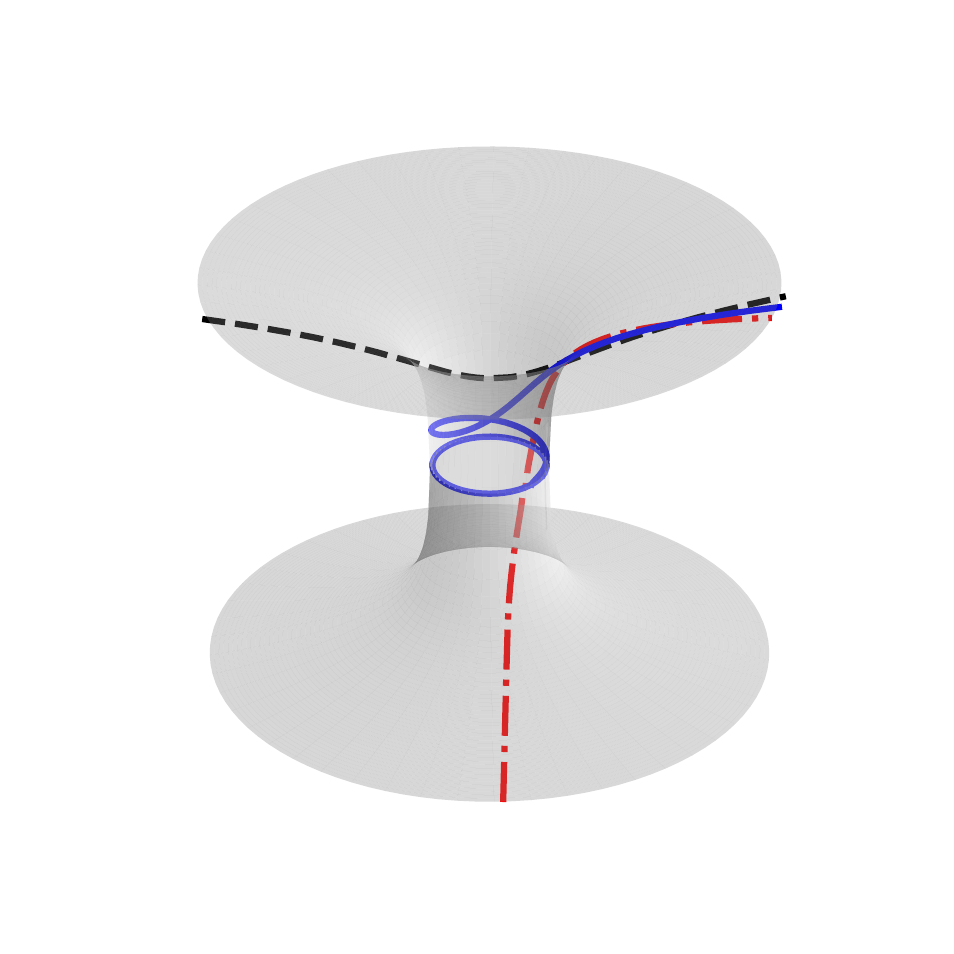}\

\includegraphics[width=0.3\textwidth]{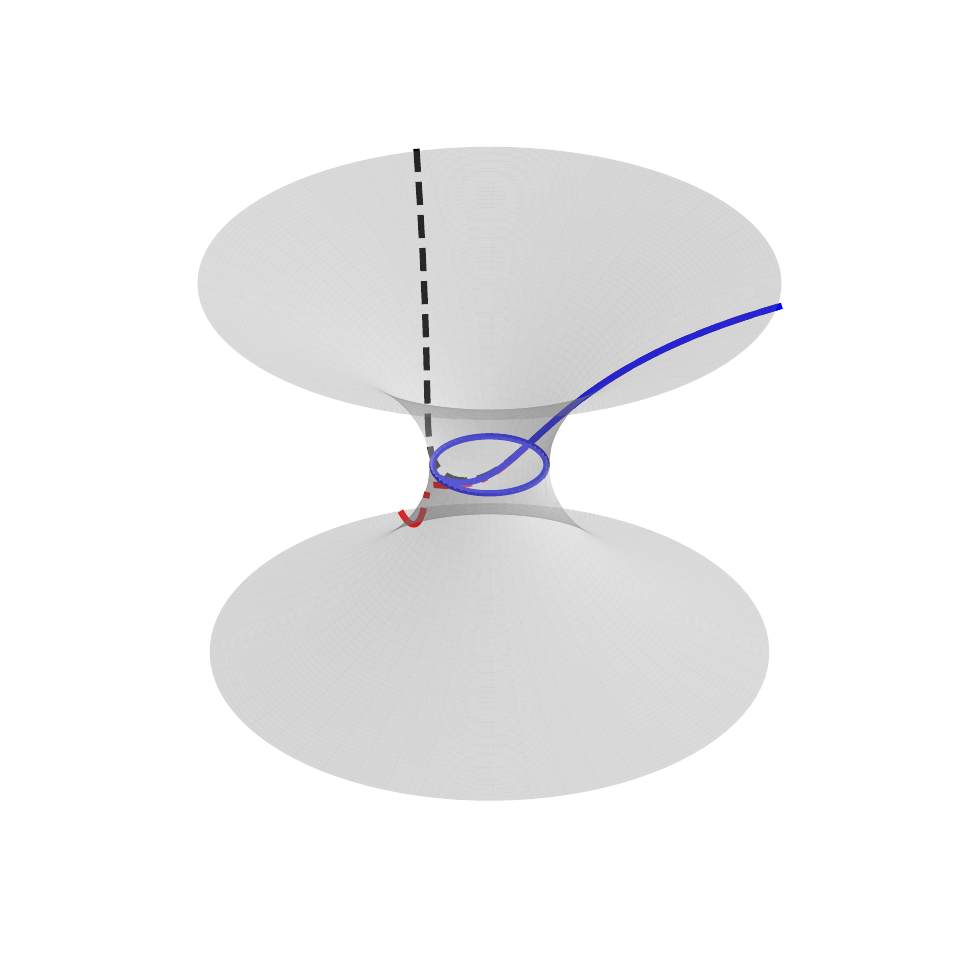}\
\includegraphics[width=0.3\textwidth]{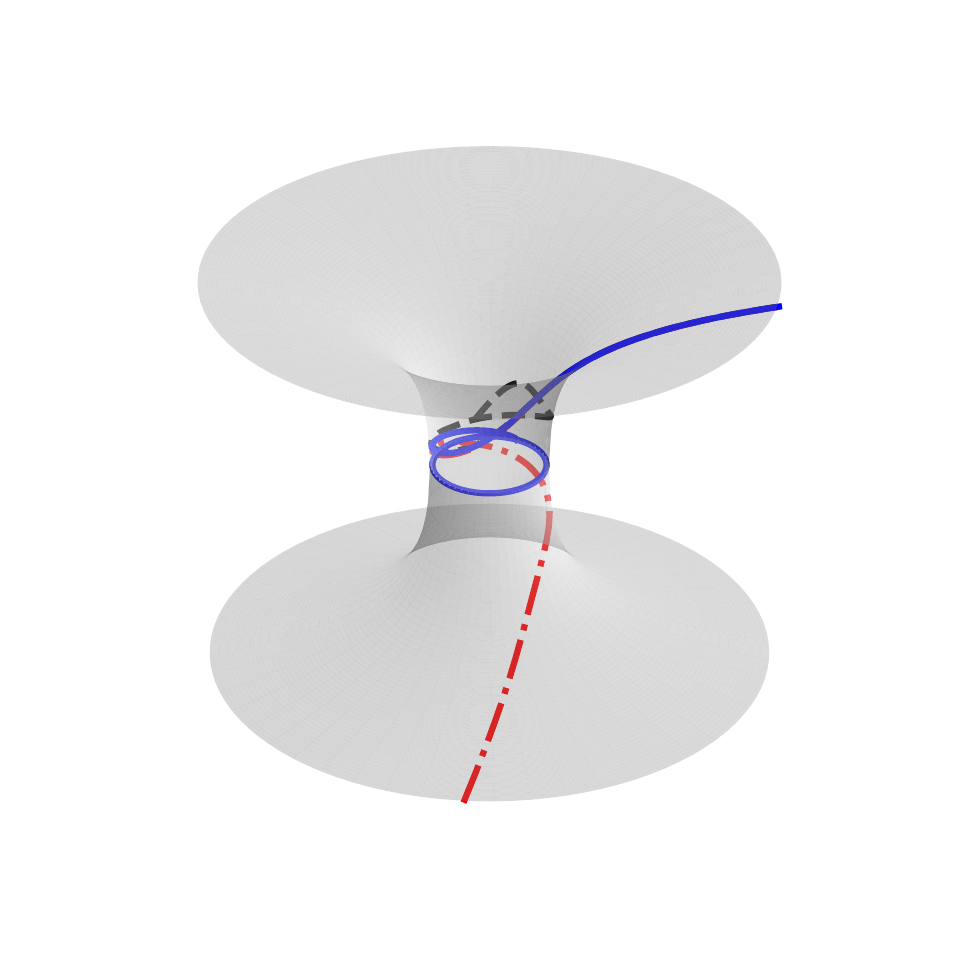}\
\includegraphics[width=0.3\textwidth]{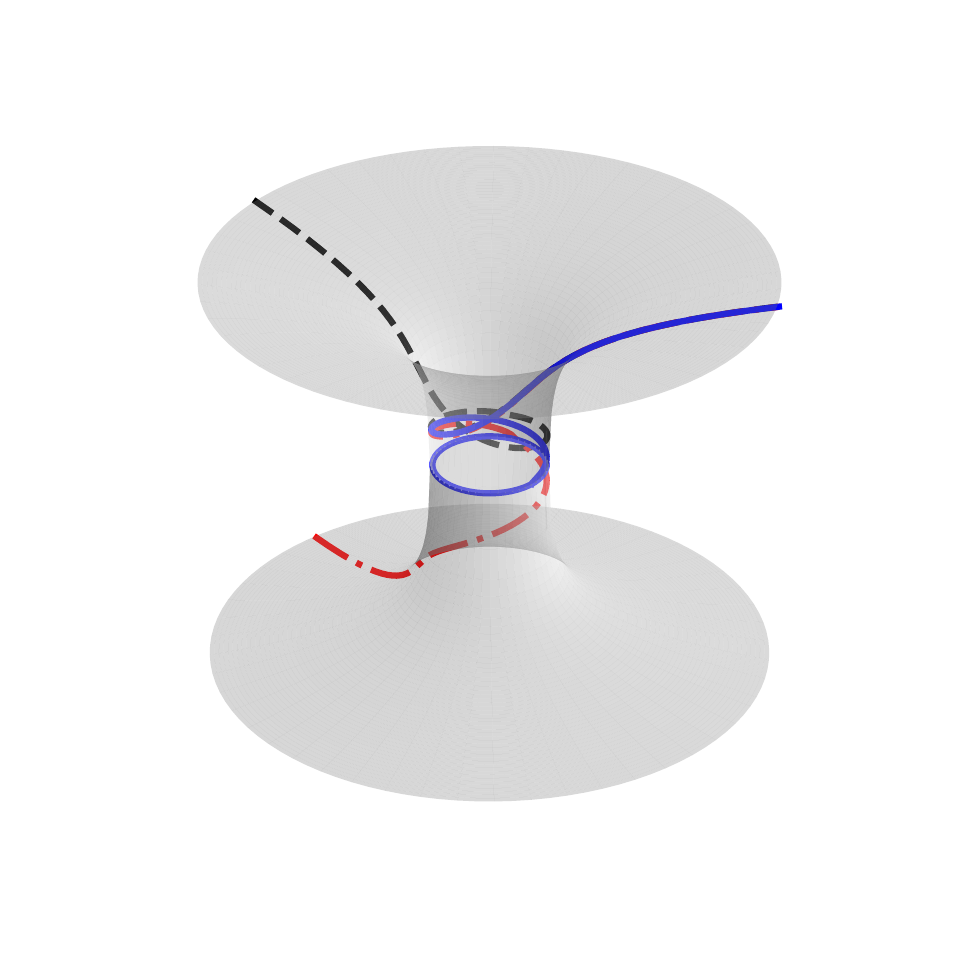}\

\includegraphics[width=0.3\textwidth]{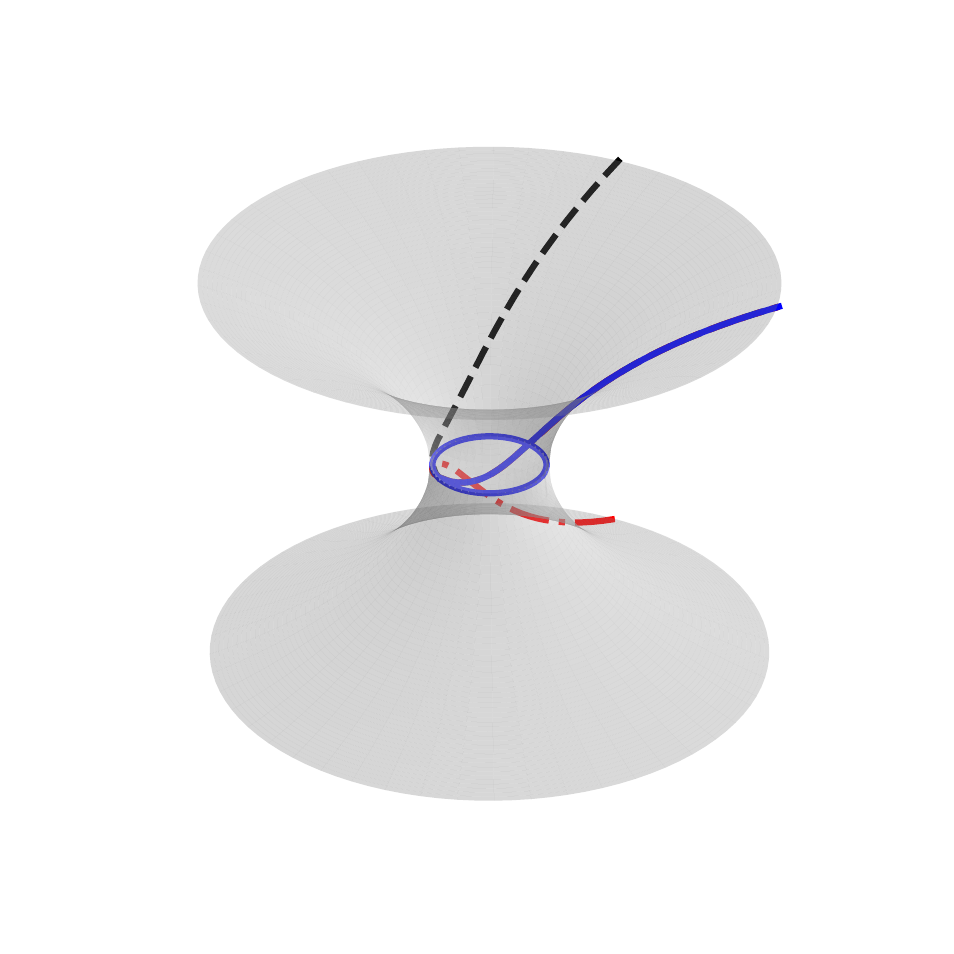}\
\includegraphics[width=0.3\textwidth]{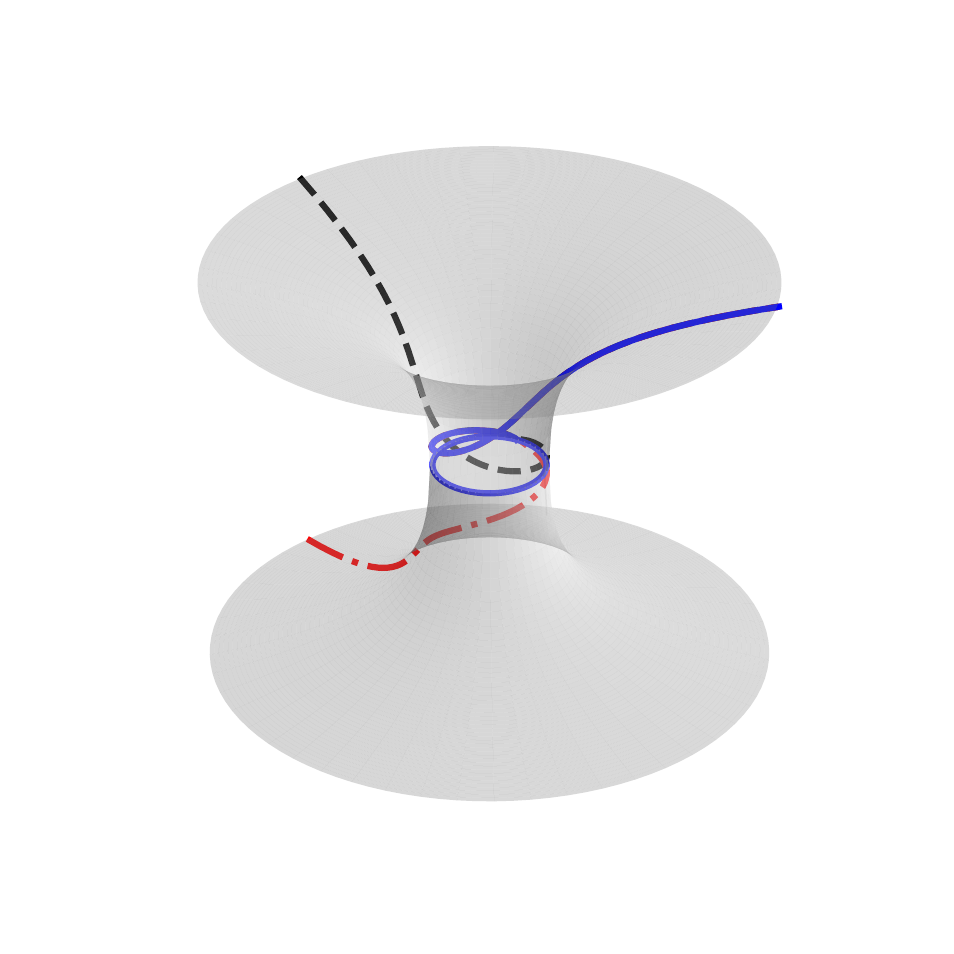}\
\includegraphics[width=0.3\textwidth]{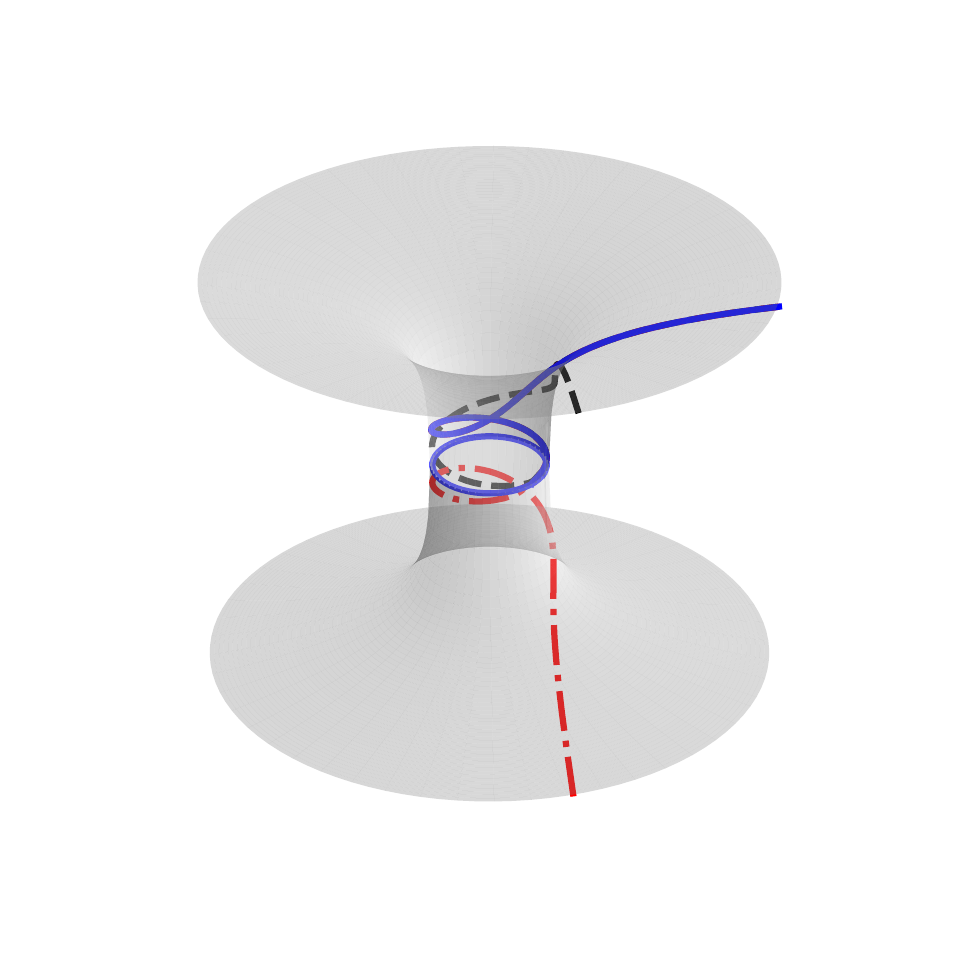}\
\caption{\label{tlo}
Time-like geodesics. This is for $c = 0.4$, $d = 0.2$ (left column), $c = 1.2$, $d = 0.6$ (center column), $c = 1.6$, $d = 1.8$ (right column). The black curve is the geodesic for the particle with $L = L_c + \delta$, the green curve is the geodesic for the particle with $L = L_c - \delta$, and the blue curve is the geodesic for the particle with $L = L_c$. This is for $\delta = 0.5$ (first row), $\delta = 0.005$ (second row), and $\delta = 0.00005$ (third row).}
\end{figure*}

For the critical case  $L = L_c$ we consider only one integral
\begin{equation}\label{geocrit}
    \phi (r) =  \int_{r_i}^{r} \frac{L}{r'^2 \sqrt{\left( 1-\frac{b(r')}{r'}\right)\left(\epsilon+E^2 -\frac{L^2}{r'^2} \right)}} dr' + \phi_i
\end{equation}
The results obtained by integrating (\ref{geo1}) and  (\ref{geo2}) numerically for timelike geodesics $\epsilon = 1$ for different wormhole parameters,  are shown in Fig.\ref{tlo} which corresponds to an embedding diagram of the wormhole. More precisely, Fig. \ref{tlo} has been obtained by setting $dt=0$ (as the solution is static), and $d\theta=0$ with $\theta=\pi/2$ (as the solution is spherically symmetric). The resulting metric is embedded in the space and is parameterized by $(r,\phi)$. Note that these results are in accordance with the previous discussion. The red line corresponds to particles traveling from one universe (upper branch of the solid of revolution) to another universe (bottom branch). In contrast, the black line represents the trajectory of a particle that reaches the throat but remains in the same universe (is always in the upper part of the solid in revolution). Finally, the blue line stands for particles that get trapped at the throat and describe circular orbits. It is worth noticing that, as we have established the same initial conditions in the numeric integration, all the trajectories overlap to the right of the upper branch of the embedding diagram. 

Now, let us discuss how the parameters $c, d, \delta$ influence the trajectories. It is noticeable that, for fixed $c$, and $d$, the particles orbit around the throat as $\delta$ decreases (approaching the critical condition) before reaching their final state. This effect is evident in the first column of the figure where $c=0.4$ and $d=0.2$, and where the black and red curves are seen to twist around the throat. A similar scenario occurs in the second and third columns where the black and red curves loop more times around the throat as the values of the pair $c, d$ increase. 

A similar analysis can be performed to determine the orbits of photons ($\epsilon = 0$). In this case, we can rewrite the equation of motion for null geodesics in terms of the impact parameter $\beta = L/E$
\begin{equation} \label{ligh rings}
    \frac{d\phi}{dr}  = \pm \frac{\beta}{r^2 \sqrt{ \left( 1-\frac{b(r)}{r}\right)\left(1-\frac{\beta^2}{r^2} \right)}}
\end{equation}
The image that receives the observer is the lensed trajectories of photons due to gravitational effects, thus it is useful to use the ray-tracing method \cite{guerrero_olmo_rubiera-garcia_gómez_2022,mishra_chakraborty_2018,taylor_2014,Olmo:2021piq} which consists in
the backtracking of the light ray that appears on the observer's view by using (\ref{ligh rings}) in order to determine where it was generated. 

For null geodesics, replacing the effective potential Eq. (\ref{effpot}) in Eq. (\ref{eqgeo}), the geodesic equation in terms of the impact parameter $\beta$ and the angular momentum $L$ reads

 \begin{align}\label{eqgeonull}
     \dot{x}^2 L^2 = \frac{1}{\beta^2}- \frac{1}{r(x)^2}.
 \end{align}
From (\ref{eqgeonull}) one can find, for each impact parameter $\beta$,  the closest point from the throat $x_{min}$ in which the photons are deflected (satisfying $\dot{x}|_{x_{min}} = 0$) such that $\frac{1}{\beta^2}=\frac{1}{r(x_{min})^2}$. The critical impact parameter $\beta_c$ corresponds to the maximum of the effective potential
\begin{equation}\label{impact}
    \frac{1}{\beta_c^2} = V(x_c),\:\; \frac{dV}{dx}\bigg|_{x_c} = 0,\:\; \frac{d^2V}{dx^2}\bigg|_{x_c} < 0,
\end{equation}
where $x_c$ is the $x_{min}$ of a photon with impact parameter $\beta_c$, i.e., $x_c$ is the closest point to the throat that a photon with critical impact parameter can be.
The photon will follow an unstable circular orbit at $r_c = r(x_c)$ (radius of the photon sphere). Equation (\ref{impact}) implies $x_c = 0$, thus $r_c = r_0$ so the photon spheres are only observed at the throat.  Note that a photon with $\beta > \beta_c$ will be deflected in the same universe while a photon with $\beta < \beta_c$ will traverse through the throat and continue in the other universe. It is worth noticing that as $\beta$ approaches to $\beta_c$ the photon follows an orbit that turns around the throat of the wormhole an increasing number of times. 

\begin{figure*}[hbt!] 
\centering
\includegraphics[width=0.45\textwidth]{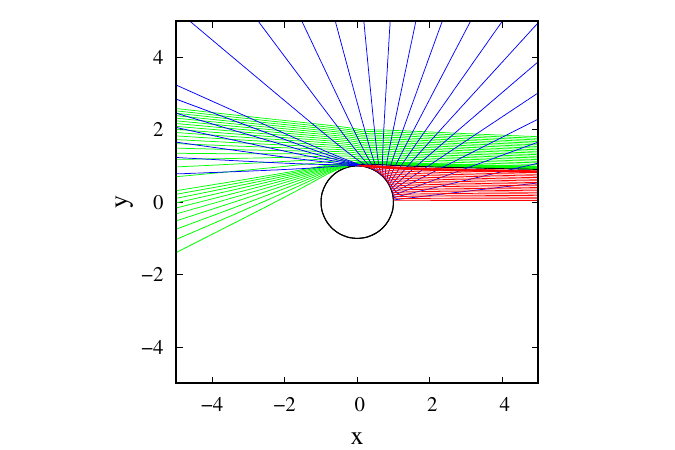}\
\includegraphics[width=0.45\textwidth]{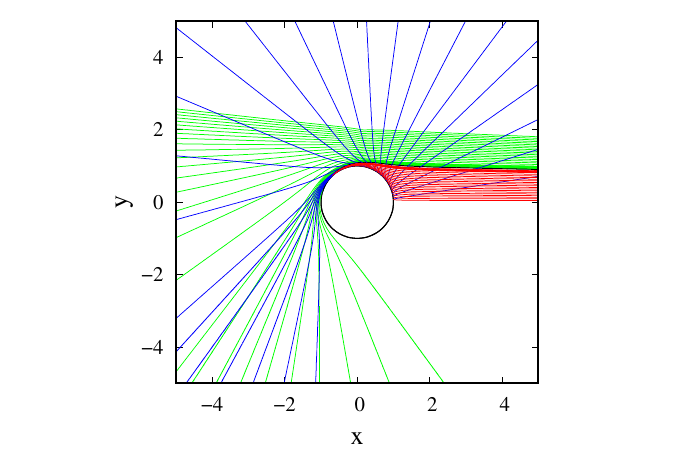}\
\includegraphics[width=0.45\textwidth]{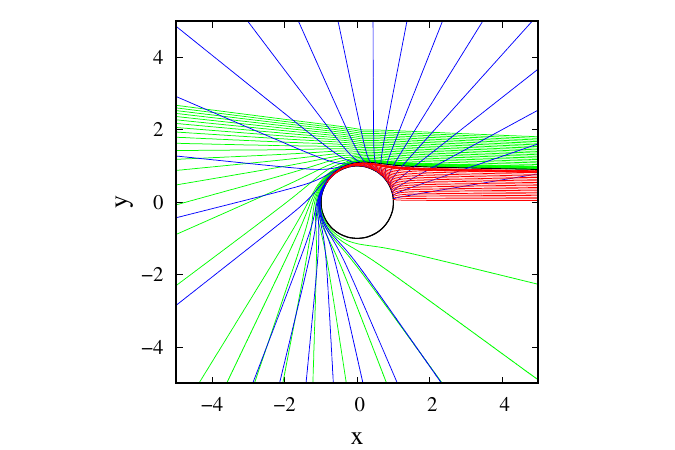}\
\caption{\label{lr}
Ray tracing  for $c = 0.4$, $d = 0.2$ (left), $c = 1.2$, $d = 0.6$ (center), and $c = 1.6$, $d = 1.8$ (right). The green curves represent incoming light rays with $\beta > \beta_c$. The red curves represent incoming light rays with $\beta < \beta_c$, and the blue curves represent outgoing light rays in the other universe.}
\end{figure*}

The images shown in Fig. \ref{lr} are the ray-tracing solutions of 
\eqref{ligh rings}. The observer is placed at the left of the screen at infinity and we see the trajectories of the photons from the north pole. By solving \eqref{ligh rings} we consider the initial conditions in Cartesian coordinates $(x_i, y_i)$, where $x_i = 100 r_0$ is the numerical infinity and $y_i = \beta$. The impact parameter is varied in the interval $[0.5,2]$ with a step of $0.05$. In the figure, the green curves represent light rays that pass close to the throat, deviate, and continue within the same universe. The red lines represent light rays that reach the throat and travel to the other universe (or an asymptotically flat region of the same universe), as represented by the blue curves. It can be seen that as the pair $c, d$ increases, the curves twist more around the wormhole's throat. We want to conclude this section with a brief comment about how we can differentiate the behavior described here from the observed in a black hole geometry. Note that, although the particles can describe similar trajectories in a black hole geometry, they cannot disappear as they reach the horizon radius. In contrast, the redshift becomes infinity and for an asymptotic observer, the particle never reaches the horizon.

\section{Accretion}\label{accretion}
In this section, we delve into the study of steady and spherically symmetric accretion on traversable wormholes. Although accretion has been a topic primarily in the context of black holes \cite{Mach:2013gia, Yang:2015sfa,Debnath:2015hea, Bahamonde:2015uwa, Ahmed:2016cuy, Azreg-Ainou:2018wjx, UmarFarooq:2019uqr, Neves:2019ywx, Panotopoulos:2021ezt, Cai:2022fdu}, the growing interest in traversable wormholes has led to an increasing interest in the accretion processes on these objects as well.

\begin{figure*}[hbt!] 
\centering
\includegraphics[width=0.3\textwidth]{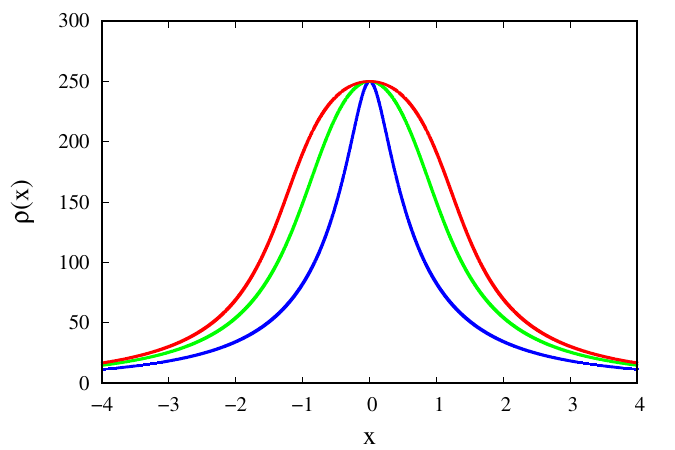}
\caption{\label{densplot}
Dust energy density as a function of the tortoise coordinate for c = 0.4, d = 0.2 (blue); c = 1.2, d = 0.6 (green); c = 1.6, d = 1.8 (red).
}
\end{figure*}

\begin{figure*}[hbt!] 
\centering
\includegraphics[width=0.3\textwidth]{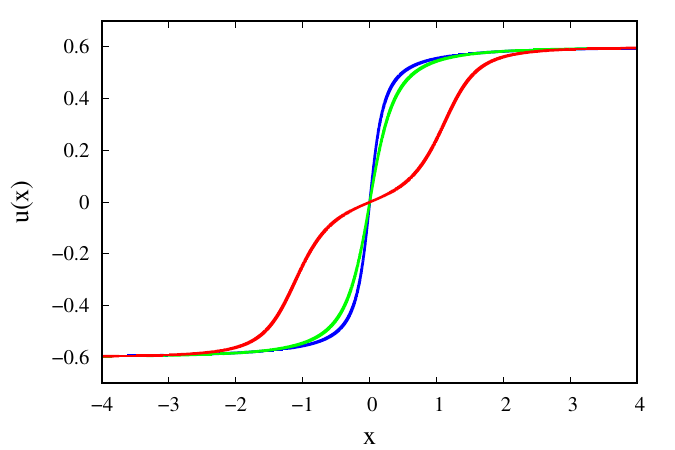}
\caption{\label{velplot}
Dust velocity as a function of the tortoise coordinate for c = 0.4, d = 0.2 (blue); c = 1.2, d = 0.6 (green); c = 1.6, d = 1.8 (red).
}
\end{figure*}

In this section, we explore the steady spherically symmetric accretion of pressureless fluid (dust) in the vicinity of traversable wormholes. In what follows, we will introduce the basic concepts we will need for future developments based on Refs. \cite{Bahamonde:2015uwa}. The energy-momentum tensor for dust is
\begin{equation}\label{tmunu}
    T_{\mu \nu}=\rho u_{\mu}u_{\nu},
\end{equation}
where $\rho$ is the energy density and $u^{\mu}$ is the 4-velocity
\begin{equation}
    u^{\mu}=\frac{dx^{\mu}}{d\tau}=(u^t,u^r,0,0)
\end{equation}
with $\tau$ the proper time. Note that the components of the 4-velocity and energy density depend only on $r$ and the angular components are zero due to the spherical symmetry we are assuming here. From the normalization condition $u^{\mu}u_{\mu}=-1$ we find
\begin{equation}
    u^t \equiv \frac{dt}{d\tau}=\sqrt{\frac{u^2}{\left(1-\frac{b(r)}{r}\right)}+1},
\end{equation}
where $u \equiv u^r $. The energy-momentum tensor conservation law $T^{\mu \nu}\,_{;\mu}=0$ implies
\begin{equation}\label{encons}
    \frac{\rho u r^2}{1-\frac{b(r)}{r}}\sqrt{u^2+\left(1-\frac{b(r)}{r}\right)}=A_1
\end{equation}
where $A_1$ is an integration constant. 
The second equation of motion for the accretion process is the mass flux conservation law $J^{\mu}\,_{;\mu}=0$ from where
\begin{equation}\label{mascons}
    \frac{\rho u r^2}{ \sqrt{1-\frac{b(r)}{r}}}=A_2
\end{equation}
where $A_2$ is an integration constant. Dividing \eqref{encons} by \eqref{mascons} we obtain a useful relation that does not depend on $\rho$
\begin{equation}\label{velo}
    \sqrt{\frac{u^2}{\left(1-\frac{b(r)}{r}\right)}+1}=\frac{A_1}{A_2} \equiv A_4,
\end{equation}
with $A_4$ a constant. It is worth noticing that Eqs. (\ref{encons}) and (\ref{mascons}) play a main role in our analysis as they allow us to obtain expressions for the radial velocity and the energy density, namely
\begin{eqnarray}
     u &=& \pm \sqrt{\left(1-\frac{b(r)}{r}\right)(A_4^2-1)}\\
    \rho &=& \frac{A_2}{\sqrt{A_4^2-1}}\frac{1}{r^2},
\end{eqnarray}
where $+(-)$ indicates ingoing (outgoing) fluid and
the constants $A_4$ and $A_2$ can be calculated from boundary conditions (far from the throat of the wormhole). In our case, we set the boundaries of the system at $r_i = 500 r_0$ where the initial radial velocity and density have been set as $u_i = 0.6$ and $\rho_i = 0.001$, respectively. The profiles for the energy density and radial velocity are shown in Figs. \ref{densplot} and \ref{velplot}, respectively.

In Fig. \ref{densplot} we note that the energy density profile corresponds to a bell-shaped curve, with its peak at the center of the throat of the wormhole. This behavior suggests that the accretion of dust is concentrated around the throat region, where the curvature of spacetime is the most pronounced. To be more precise, the bell-shaped distribution indicates that as we move away from the throat in both directions along the tortoise coordinate, the energy density of the accreting dust gradually decreases. This behavior partially coincides with what occurs for dust accretion on a Schwarzschild de Sitter black hole with a topological defect reported in \cite{Bahamonde:2015uwa}. The main difference is that, in the black hole case, the density diverges as we approach to the horizon.

In Fig. \ref{velplot} we note that the radial velocity of the dust decreases as we approach the throat of the wormhole, ultimately reaching zero at the throat itself. This behavior is the opposite of the observed in the case of black holes where the velocity increases instead. 

Another important quantity that deserves to be explored is the rate of change of the mass $\dot{M}$ of the dust that traverses the wormhole which corresponds to the integral of the flux of the fluid over the 2-sphere, namely \cite{Debnath:2015hea}
\begin{equation}
    \dot{M}=\int T_0^1 \sqrt{-g} d\theta d\phi. 
\end{equation}
Combining (\ref{metric}) and (\ref{tmunu}), the integration leads to
\begin{equation}\label{mass}
     \dot{M}=4\pi r^2 \rho u \sqrt{\frac{u^2}{\left(1-\frac{b(r)}{r}\right)}+1},
\end{equation}
from where 
\begin{equation}\label{mass2}
     \dot{M}=4\pi A_2\left(1-\frac{b(r)}{r}\right)\sqrt{u^2 \left (1-\frac{b(r)}{r} \right)+1}
\end{equation}
It should be emphasized that in contrast to what occurs in the case of black holes, the accreted fluid is not ``accumulated'' by the wormhole but the fluid ``travels'' between universes connected by the wormhole throat. In the case of black holes, Eq. (\ref{mass2}) indicates that the mass of the black hole must increase. To be more precise, $\dot{M}$ is not interpreted as the ``mass gained'' by the wormhole but the flux of particles traversing the throat to another universe.

\begin{figure*}[hbt!] 
\centering
\includegraphics[width=0.3\textwidth]{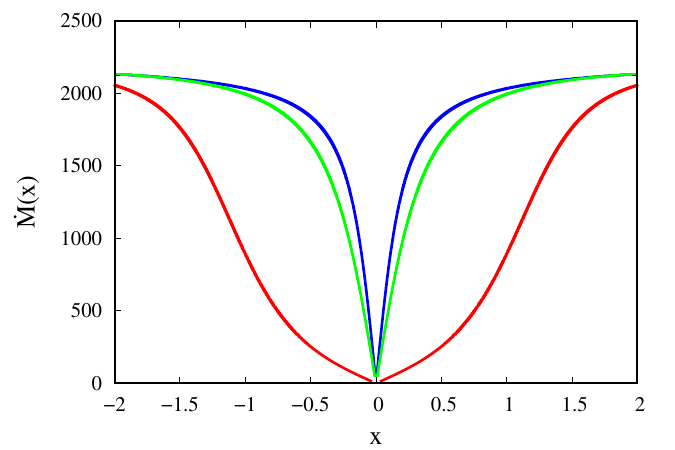}
\caption{\label{massplot}
Rate of change of the mass as a function of the tortoise coordinate for c = 0.4, d = 0.2 (blue); c = 1.2, d = 0.6 (green); c = 1.6, d = 1.8 (red).
}
\end{figure*}

In Fig. \ref{massplot} we plot $\dot{M}$ for different parameters of the wormhole. The plot reveals that the rate of mass change decreases as we approach the throat of the wormhole as occurs with the reduction in radial velocity. 


\section{conclusions}
In this work, we have studied the behavior of massive and massless particles in the vicinity of a traversable wormhole. In particular, we have studied their geodesic trajectories and stationary accretion of dust. We have found that there are two types of open orbits: those corresponding to particles deflected by the throat of the wormhole and remaining in the same universe, and those that traverse the throat into a new universe or a region asymptotically far from the same universe. There is also an extreme case corresponding to unstable particle orbits around the throat. In all three cases, it is observed that the wormhole's free parameters strongly influence the shape of the particles. To be more precise, for certain values, the open orbits twist around the throat before exiting. Regarding accretion, we have found several interesting results compared to what is found for black holes. In particular, all the quantities studied, such as dust density, radial velocity, and the rate of change of mass crossing the wormhole, are finite in the entire spacetime, unlike what occurs in the case of black holes where divergences are found near the event horizon.

Before concluding this section,
 a couple of comments are in order. First, it should be interesting to explore the steady spherical accretion for more complex fluids as those fulfilling a polytropic equation of state, for example. Polytropic accretion has been broadly explored in the context of black holes (see, \cite{1,2,3} for example) and it could be interesting to explore to what extent the process differs when we consider a traversable wormhole geometry.

\end{document}